\documentstyle[aps,prl]{revtex}
\begin{document}
\draft
\twocolumn[\hsize\textwidth\columnwidth\hsize\csname %
@twocolumnfalse\endcsname

\title{ Spectral properties of the planar $t-J$ model}
\author{ J. Jakli\v c$^1$ and P. Prelov\v sek$^{1,2}$ }
\address{ $^{1}$J. Stefan Institute, University of Ljubljana, 1001
Ljubljana, Slovenia }
\address{$^{2}$Institut Romand de Recherche Num\'erique en Physique des
Mat\'eriaux (IRRMA), PHB-Ecublens, CH-1015 Lausanne, Switzerland}
\date{\today}
\maketitle
\begin{abstract}\widetext
The single-particle spectral functions $A({\bf k},\omega)$ and
self-energies $\Sigma({\bf k},\omega)$ are calculated within the $t-J$
model using the finite-temperature Lanczos method for small systems. A
remarkable asymmetry between the electron and hole part is found. The
hole (photoemission) spectra are overdamped, with ${\rm Im} \Sigma
\propto \omega$ in a wide energy range, consistent with the marginal
Fermi liquid scenario, and in good agreement with experiments on
cuprates. In contrast, the quasiparticles in the electron part of the
spectrum show weak damping.
\end{abstract}
]
\narrowtext
The normal state of superconducting cuprates in many aspects
contradicts the phenomenology of the normal Fermi liquid
(FL). Anomalous frequency and temperature dependence of several
response functions is generally attributed to
electronic correlations, yet a proper description is missing so far.
The angle resolved photoemission (ARPES) experiments
\cite{shen,ding,olson} probe the one-particle spectral
function $A({\bf k},\omega)$.  At intermediate doping they reveal for
a wide class of cuprates a well defined large Fermi surface (FS)
consistent with the Luttinger theorem and similar quasiparticle (QP)
dispersion \cite{ding}. This seems to imply the validity of the
concept of the usual metal with electronic-like FS.  Such simple FL
picture is in an apparent contradiction with magnetic and transport
properties, e.g. electrical conductivity scales with hole
concentration, closer to the picture of holes moving in the
antiferromagnetic (AFM) background.  Moreover, in 
ARPES the FL interpretation is spoiled by the overdamped character
of QP peaks \cite{olson,ding}. Although a large background makes
fits of particular lineshapes non-unique \cite{olson,liu}, the QP
inverse lifetime is found to be of the order of the QP energy,
i.e. $\tau^{-1}\propto \omega$ for $\omega>T$, leading to the concept
of the marginal Fermi liquid (MFL) \cite{varma} with an anomalous
single-particle and transport relaxation, in contrast to
$\tau^{-1}\propto\omega^2$ in the normal FL.

It is unclear whether above features can be reproduced within generic
models of strongly correlated systems, such as the Hubbard and the
$t-J$ model, in particular in the most challenging regime of
intermediate doping.  Spectral properties of these 2D models have been
so far studied mainly via numerical techniques \cite{dagorev},
e.g. exact diagonalization (ED) \cite{stephan} and Quantum Monte Carlo
(QMC) \cite{bulut}.  These studies, as well as some analytical
approaches \cite{wang}, established a reasonable consistency of the
model QP dispersion with the experimental one, as well as the
possibility of large FS, but have not been able to investigate closer
the character of QP, being in the core of the anomalous low-energy
properties.

The aim of the present work is to employ the finite-temperature
Lanczos method \cite{jplanc} to calculate $A({\bf k},\omega)$ within
the $t-J$ model. This method has been already applied to other dynamic
\cite{jpdyna} and static \cite{jpterm} functions, yielding features
consistent with the MFL concept and experiments on cuprates. Although
calculations are still done in small systems, by using finite (but
small) $T>0$ smooth enough spectra are obtained not only to determine
the QP dispersion, but for the first time also the spectral lineshapes
and corresponding self-energies.

We study the $t-J$ model \cite{rice}
\begin{equation}
H=-t\sum_{\langle ij\rangle  s}(\tilde{c}^\dagger_{js}\tilde{c}_{is}+
\text{H.c.})+J\sum_{\langle ij\rangle} ({\bf S}_i\cdot {\bf S}_j -
{1\over 4} n_i n_j) \label{model}
\end{equation}
on the planar square lattice and set $J/t=0.3$ to address the regime
of cuprates.  The operators $\tilde{c}_{js},\tilde{c}^\dagger_{js}$
project out the states with doubly occupied sites.  The spectral
properties of the model Eq.~(\ref{model}) are investigated by
calculating the retarded Green's function ($\mu$ is chemical potential)
\begin{equation}
G({\bf k},\omega)=-i\int_0^\infty dt\;e^{i(\omega+\mu)t}
\langle\{\tilde{c}_{{\bf k}s}(t),\tilde{c}^\dagger_{{\bf k}s}(0)\}
\rangle.
\label{green}
\end{equation}
The average is grandcanonical,
which in actual calculations at low $T$ in a system with $N$ sites and
fixed hole concentration $c_h=N_h/N$ is replaced by a canonical one in
the subspace of states with $N_h$ holes.  The two anticommutator terms
correspond at low $T$ to the hole -- inverse photoemission spectra
(IPES), and the electron -- photoemission spectra (PES), respectively.

The calculation of $G({\bf k},\omega)$ at $T=0$ with the ED technique
is well established \cite{stephan,dagorev}, but a small number of
sharp peaks in the spectra makes it difficult to extract information
on lineshapes and self energies.  The QMC methods resort to the use of 
maximum entropy analysis \cite{bulut}, which also leads to quite 
restricted $\omega$-resolution.  The $T>0$ Lanczos
method \cite{jplanc} eliminates these problems for dynamic quantities,
i.e.  yields smoother spectra and allows for study of the
$T$-dependence. The requirement is however that $T>T_{\rm fs}$, where
$T_{\rm fs}$ is the characteristic temperature at which in a given
small system the finite-size effects set in (for discussion of the
method we refer to previous works \cite{jplanc,jpdyna}). We have
calculated the Green's function Eq.~(\ref{green}) on systems with N=16
and 18 sites using $\sim 120$ Lanczos steps and sampling over $\sim
1000$ random states. The finite-size effects are small at $T\agt
T_{\rm fs}(N,N_h)$, where, e.g., $T_{\rm fs}\sim 0.1t$ for
$N_h/N=3/16$.

From the Green's function we obtain the spectral function $A({\bf
k},\omega)=-(1/\pi){\rm Im}G({\bf k},\omega)$ and the one-particle
density of states (DOS) ${\cal N}(\varepsilon)=(2/N)\sum_{\bf k}
A({\bf k},\varepsilon-\mu)$.  The latter is used to define the zero of
energy and thus the chemical potential in Eq.~(\ref{green}) via
$\int_{-\infty}^\infty {\cal N}(\omega+\mu)
(e^{\beta\omega}+1)^{-1}d\omega=1-c_h$.  We find a very good agreement
between $\mu$ calculated this way and from the thermodynamic function
$c_h(T,\mu)=N_h/N$
\cite{jpterm}.

Of particular interest is the self-energy
\begin{equation}
\Sigma({\bf k},\omega)=\omega-G({\bf k},\omega)^{-1}.
\end{equation}
The relation contains no free term, in contrast to the usual
definition, since the $t-J$ model does not allow for a free-fermion
propagation even at $J=0$.  It is also important to note that due to
projected fermion operators in the model the spectral function $A({\bf
k},\omega)$ is not normalized to unity \cite{stephan}, but rather to
$\langle\{\tilde{c}_{{\bf k}s},\tilde{c}^\dagger_{{\bf k}s}\}\rangle =
(1+c_h)/2$. This has several consequences, e.g.  ${\rm Re} \Sigma({\bf
k},\omega\to \infty)$ does not vanish, but varies linearly with
$\omega$.

In Fig.~\ref{fig1} we first present $ A({\bf k},\omega)$ for systems
with $c_h \sim 0.12$ (combining results for $N_h=2$ on systems with
$N=16,18$) and $c_h=3/16$. The spectra are broadened to Lorentzians
of variable width
$\delta=\delta_0+(\delta_\infty-\delta_0)\tanh^2(\omega/\Delta)$, with
$\delta_\infty =0.2t$, $\delta_0=0.04t$, and $\Delta=1.0t$.  In
this way sharper (well resolved) low-energy features remain
unaffected, while the fluctuations at higher $\omega$, mainly due to
restricted sampling, are smoothened out. In any case, $\delta$ is
always smaller than the energy scale of main spectral features.

We observe in Fig.~\ref{fig1}, presented at all available ${\bf k}$, a
coexistence of sharper features, associated with coherent QP peaks,
and of a pronounced incoherent background, as already established in
earlier studies \cite{stephan}. The coherent peaks in Fig.~\ref{fig1}
disperse through $\omega=0$ as ${\bf k}$ crosses the FS. Within the
given resolution in the ${\bf k}$-space the FS appears to be large
already for $c_h=2/18$, consistent with the Luttinger theorem. The
total QP dispersion $W$ is broadened as $c_h$ is increased,
qualitatively consistent with the slave boson picture where $W
\propto c_h t + \chi J$ \cite{wang}.

In Fig.~\ref{fig2} we show $\Sigma({\bf k},\omega)$ at $c_h=3/16$ and
at lowest $T=0.1t\sim T_{\rm fs}$. We first notice an asymmetry
between the PES ($\omega<0$) and IPES ($\omega>0)$ spectra at all
${\bf k}$. ${\rm Im}\Sigma$ are small for $\omega>0$, as compared to
$\omega<0$. For ${\bf k}$ outside FS this implies a weak QP damping,
consistent with sharp IPES peaks seen in $A({\bf k},\omega)$,
Fig.~\ref{fig1}, containing the major part of the spectral
weight. ${\rm Re} \Sigma$ shows an analogous asymmetry, in the region
$\omega>0$ resembling moderately renormalized QP. Due to projections
in Eq.~(\ref{model}), the slope in ${\rm Re}\Sigma$ is not zero even
at $|\omega|\gg t,J$.

The behavior on the PES ($\omega<0$) side is very different.  For all
${\bf k}$, ${\rm Im}\Sigma$ are very large (several $t$ away from
$\omega \sim 0$), leading to overdamped QP structures.  We should here
distinguish two cases. For $\bf k$ well outside FS, ${\rm Im}\Sigma>t$
does not invalidate a well defined QP (at $\omega>0$), but rather
induces a weaker reflection (shadow) of the peak at $\omega <0$, as
well seen in Fig.~\ref{fig1} for ${\bf k} =(\pi,\pi)$.  On the other
hand, the $\omega$ variation for ${\bf k}$ inside or near the FS is
more regular, and can be directly related to the QP damping.
Particularly remarkable feature, found in Fig.~\ref{fig2}, is a linear
frequency dependence of ${\rm Im}\Sigma$ at $\omega<0$ for ${\bf k}=
(\pi/2,0),(\pi/2,\pi/2)$.  Meanwhile ${\bf k}=(0,0)$, being further away
from the FS, seems to follow a different (more FL-type) behavior.
Such general behavior remains similar also for the lower doping
$c_h=2/18$.

To address the latter point in more detail, we show in Fig.~\ref{fig3}
the $T$-variation of ${\rm Im}\Sigma$ for both dopings at selected
${\bf k}$ below the FS. For $c_h=3/16$ the linearity of ${\rm
Im}\Sigma(\omega)$ is seen in a broad range $-2t\alt \omega
\alt 0$ at the lowest $T$ shown. Moreover, for this higher (`optimum') 
doping the $T$-dependence is close to a linear one, assuming a
small residual (finite-size) damping $\eta_0$ at $\omega=0$.  Data can
be well described by ${\rm Im}\Sigma=\eta_0+\gamma(|\omega|+\xi T)$,
with $\gamma\sim 1.4$ and $\xi\sim 3.5$, baring a similarity to the
MFL ansatz \cite{varma}, as well as to the conductivity relaxation
$\tau_c^{-1}$ found in the $t-J$ model \cite{jpdyna}. In contrast,
the $T$-dependence for $c_h=2/18$ seems somewhat different, and ${\rm
Im}\Sigma\propto
\omega$ only in the interval $-t\alt\omega\alt T$. This would indicate
the consistency with the alternative MFL form
\cite{varma}, however we should be aware that in this `underdoped'
regime finite-size effects are larger at fixed $T$.

Here we should comment on the manifestation of the FS in small
correlated systems. At $T,\omega\sim 0$ we are dealing in the
evaluation of Eq.(\ref{green}) with the transition between ground
states of systems with $N_h$ and $N_h'=N_h \pm 1$ holes,
respectively. Since these states have definite momenta ${\bf k}_0$,
they induce strong QP peaks for particular ${\bf k}={\bf k}'_0-{\bf
k}_0$ (defining in this way for a small system the FS, apparently
satisfying the Luttinger theorem), with ${\rm Im}\Sigma({\bf k},\omega
\sim 0)\sim 0$. However, the calculated $T$-variation is for a given
system meaningful only at $T>T_{\rm fs}$.

From $\Sigma({\bf k},\omega)$ we can calculate QP parameters: 
the dispersion $E_{\bf k}$, the weight $Z_{\bf k}$
and the damping $\Gamma_{\bf k}$,
\begin{eqnarray}
 E_{\bf k}&=&{\rm Re}\Sigma({\bf k}, E_{\bf k}),
\label{dispers}\\
 Z_{\bf k}&=&[1-\partial{\rm Re}\Sigma({\bf k},\omega
)/\partial\omega]_{\omega=E_{\bf k}}^{-1},\\
\Gamma_{\bf k}&=&Z_{\bf k}|{\rm Im}\Sigma({\bf k},E_{\bf k})|,
\end{eqnarray}
which are listed in Table~\ref{table1}.  We note that parameters are
of a limited meaning for $\bf k$ inside FS due to large $\Gamma$. In
particular, $E_{\bf k}$ (as well as $Z_{\bf k}$ and $\Gamma_{\bf k}$)
for ${\bf k}=(0,0)$ do not correspond to a weak QP peak at $\omega
\sim -t$, being overwhelmed by the incoherent background.  Otherwise,
the enhancement of the dispersion with $c_h$ is seen, accompanied by a
decrease of $\Gamma$ for $|{\bf k}|>k_F$.  To establish the relation
with the FL theory one has to evaluate QP parameters at the FS, ${\bf
k} ={\bf k}_F$. Of particular importance is the renormalization factor
$\tilde Z= Z_{{\bf k}_F}$.  $\tilde Z$ is still decreasing as $T$ is
lowered. Nevertheless we find a weak variation (cca. 20\%) within the
interval $0.1<T/t<0.3$, not inconsistent with the MFL form, leading to
$\tilde Z^{-1}\sim \ln(\omega_c/T)$. Regarding the size of $\tilde
Z$ (at low but finite $T>0$) we note, that the value of the momentum
distribution function $\bar n_{{\bf k}s}$ is very close to the maximum
for the $t-J$ model, $\bar n_{{\bf k}s}\sim (1+c_h)/2$, for all $|{\bf
k}|<k_F$ \cite{stephan}.  Taking the FS volume according to Luttinger
theorem and assuming that $\bar n_{{\bf k}s}$ falls monotonously with
$|{\bf k}|$, this implies the discontinuity $\tilde Z=
\delta \bar n_{{\bf k}s} <2c_h/(1+c_h)$.  We indeed find a consistent
result $\tilde Z= 0.28$ for $c_h=3/16$, while for $c_h=2/18$ the
value is still larger, possibly due to too high $T$.

An analogous argument can be used to explain the electron-hole
asymmetry of $A({\bf k},\omega)$.  Holes added to the system at $|{\bf
k}|<k_F,\omega<0$ move in an extremely correlated system, strongly
coupled to the spin dynamics \cite{prelovsek}, also following the
anomalous low-$\omega$ behavior
\cite{jpdyna}. On the other hand, states for $|{\bf k}|>k_F$
are not fully populated, allowing for a moderately damped
motion of added electrons for $\omega>0$.

Another feature is seen predominantly at smaller doping $c_h \sim
0.12$ for $|{\bf k}|>k_F$: along with the principal peak at $\omega>0$
a weak bump in the $\omega<0$ part of the spectrum appears when the
FS is crossed along $\overline{\Gamma M}$. In ${\rm Re}
\Sigma$ for ${\bf k}=(\pi,\pi)$ it emerges even as a strong oscillation,
leading to a double solution in Eq.~(\ref{dispers}) \cite{kampf}. In
ARPES this should be seen as the reappearance of the `shadow' QP
band for ${\bf k}$ above the FS, in accordance with experiments
\cite{shen} and some previous studies \cite{kampf,stephan}.  
The effect is less pronounced at larger doping $c_h=3/16$, probably
due to the reduction of the AFM correlation length.

Finally we show in Fig.~\ref{fig4} the variation of the DOS ${\cal
N}(\varepsilon)$ with doping. For a hole injected in the weakly doped
system ($c_h\sim 0.06$), a QP coherent peak (of width $\sim 2J$) is
seen at $\varepsilon \alt \mu$. Besides, a broad background (due to
well understood incoherent hole motion) is dominating lower
$\varepsilon$. At such low doping the electron part of DOS is weaker,
with the total intensity $2 c_h$ as compared to $1-c_h$ of the hole
part. With increasing $c_h$ the hole background doesn't reduce in
intensity, while the coherent peak near the Fermi energy widens and
its spectral weight reduces, reflecting the broadening of the QP
dispersion.  At the same time, the electron part of DOS is increasing,
both in the weight and in the width. Note that oscillations for
$\varepsilon>\mu$ appear in this regime due to the underdamped QP and
a restricted number of finite-size $\bf k$.

Here we mention the relation with the entropy $s$ \cite{jpterm},
assuming the low-$T$ form as follows from the FL theory \cite{agd},
i.e. $s = \pi^2 T {\cal N}(\mu)/3\tilde Z$. With ${\cal N}(\mu)$ in
Fig.~\ref{fig4}, weakly doping dependent at intermediate $c_h$ (also
quite close to the free-fermion value), and $\tilde Z \sim 0.28$ for
$c_h = 3/16$, we get $s \sim 0.29~k_B/{\rm site}$ at $T=0.1t$,
consistent with static calculations \cite{jpterm}. Nevertheless, one
should keep in mind that such $s$ represents a large increase over the
undoped (AFM) system, taking into account very few mobile holes
introduced into the system.

In the end we comment on the relevance of our results to the
understanding of the ARPES spectra in cuprates \cite{shen}. For
$\omega<0$ we notice the importance of the incoherent
background, consistent with the observation that in fitting the
experiments to either FL or MFL form an anomalously large background
must be assumed \cite{liu}.  For $|{\bf k}|<k_F$ we find the linewidth
typically $\Gamma\sim t$ (see Table~\ref{table1}), well compatible ($t
\sim 0.4~{\rm eV}$ in cuprates \cite{hybertsen,rice}) 
with experiments at ${\bf k}$ away from the FS and at intermediate
doping \cite{shen,ding}. Also the MFL form has been claimed to
describe better the experiments \cite{olson}, although this point is
not yet clarified \cite{shen}.  We note also that our QP dispersion
and the shape of the FS are not entirely of the form found
experimentally \cite{shen}.  This could be possibly remedied by
including the n.n.n. ($t'$) hopping term \cite{hybertsen}.  Still we
do not expect such corrections to modify conclusions
concerning the spectral shapes and the QP character.

One of the authors (P.P.) wishes to thank P. Horsch, G. Khalliulin,
R. Zeyher, and T.M. Rice for useful suggestions and fruitful
discussions, and acknowledges the support of the MPI f\"ur
Festk\"orperphysik, Stuttgart, where a part of this work has been
performed.

\begin{figure}
\caption{ Spectral functions  for $N_h=2$ holes on
systems with $N=16,18$ sites, and for $N_h=3$ holes on $N=16$ sites.}
\label{fig1}
\end{figure}

\begin{figure}
\caption{ Self energy for $c_h=3/16$ at various ${\bf k}$.}
\label{fig2}
\end{figure}

\begin{figure}
\caption{ ${\rm Im} \Sigma ({\bf k},\omega)$ for various $T$
 for the systems with $c_h=2/18$ and $c_h=3/16$ at selected ${\bf k}$
below the FS.}
\label{fig3}
\end{figure}

\begin{figure}
\caption{ Single-particle DOS ${\cal N(\varepsilon)}$
for variously doped systems. For $c_h \sim 0.06$ and
$c_h \sim 0.12$ we present joined densities for $N=16,18$ systems
with $N_h=1$ and $N_h=2$, respectively. The thin
vertical lines denote the Fermi energy $\mu$.}
\label{fig4}
\end{figure}

\begin{table}
\caption{QP parameters for two hole concentrations.}
\label{table1}
\begin{tabular}{cccccccc}
\multicolumn{4}{c}{$c_h=2/18$, $T/t=0.15$ } &
\multicolumn{4}{c}{$c_h=3/16$, $T/t=0.15$ } \\
${\bf k}$ & $E_{\bf k}/t$ & $Z_{\bf k}$ & $\Gamma_{\bf k}/t$ &
${\bf k}$ & $E_{\bf k}/t$ & $Z_{\bf k}$ & $\Gamma_{\bf k}/t$ \\ \hline
$(0,0)$          &-3.8&0.80&1.9 &$(0,0)$        &-4.2&0.73&1.4\\
$(\pi/3,\pi/3)$  &-0.7&0.26&0.65&$(\pi/2,0)$    &-1.1&0.68&1.7\\
$(2\pi/3,0)$     &-0.4&0.35&0.51&$(\pi/2,\pi/2)$& 0.0&0.28&0.32\\
$(2\pi/3,2\pi/3)$& 0.5&0.35&0.49&$(\pi,0)$      & 0.0&0.28&0.32\\
$(\pi,\pi/3)$    & 0.1&0.26&0.40&$(\pi,\pi/2)$  & 0.8&0.44&0.31\\
$(\pi,\pi)$      & 1.1&0.37&0.54&$(\pi,\pi)$    & 1.7&0.46&0.35\\
\end{tabular}
\end{table}
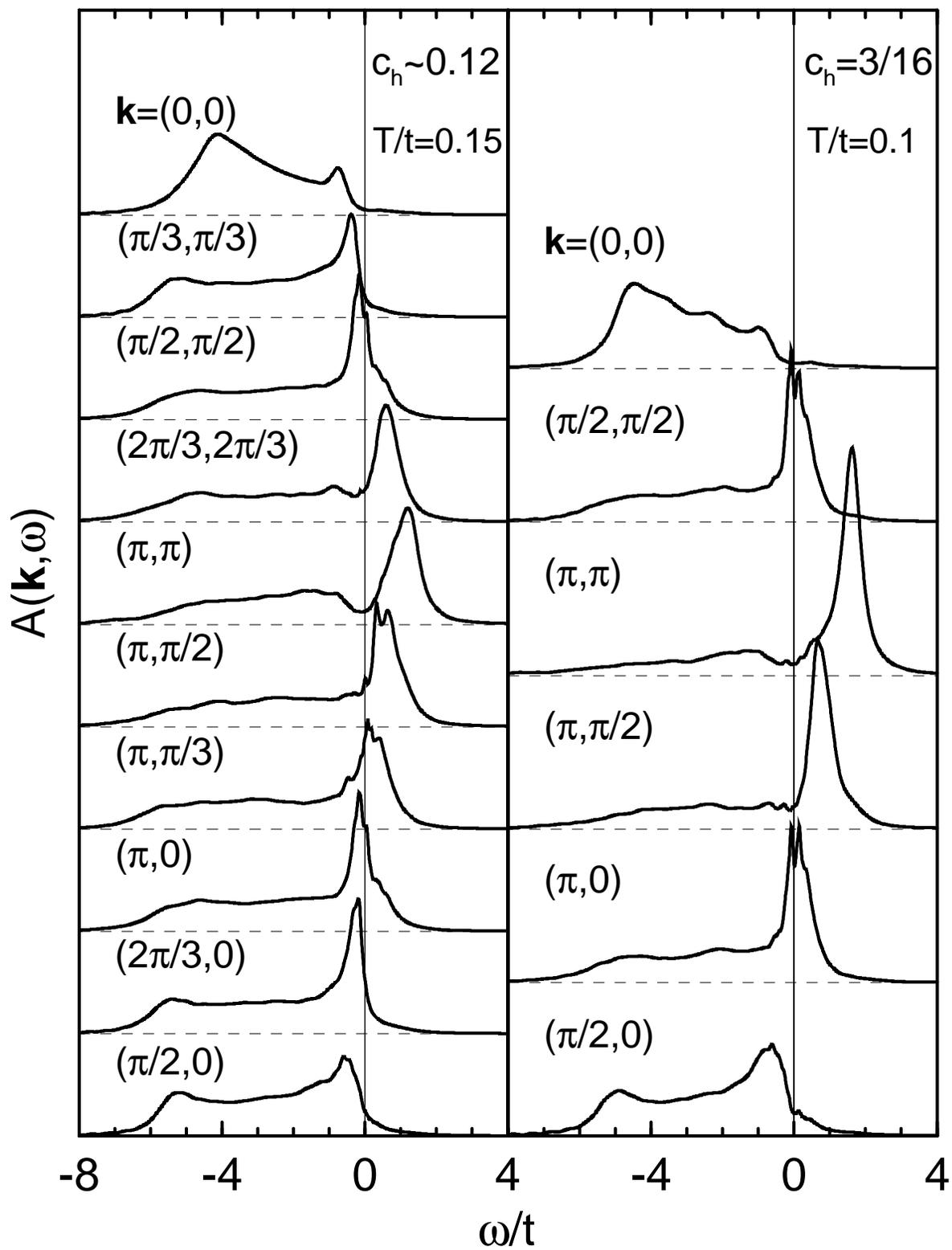
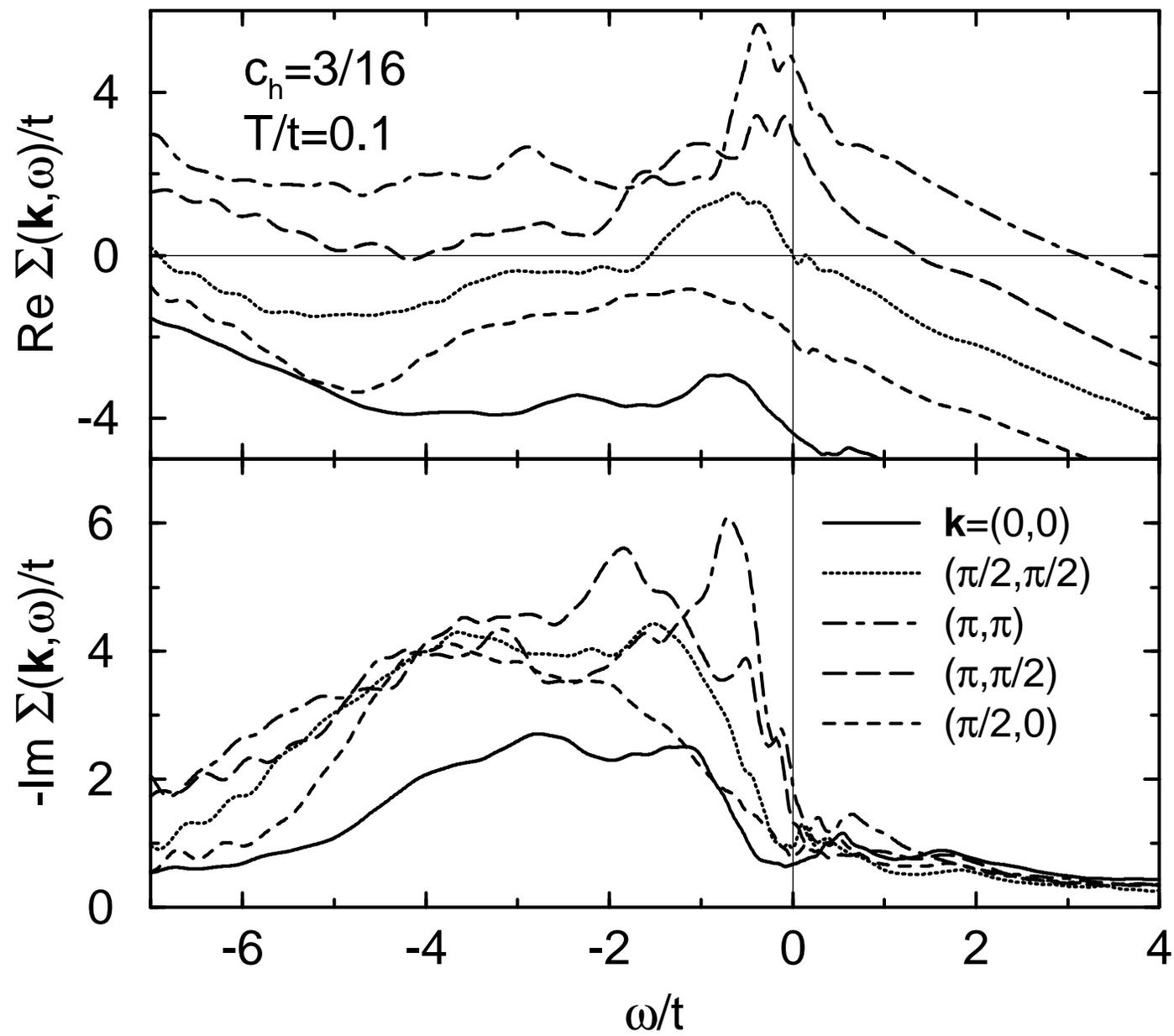
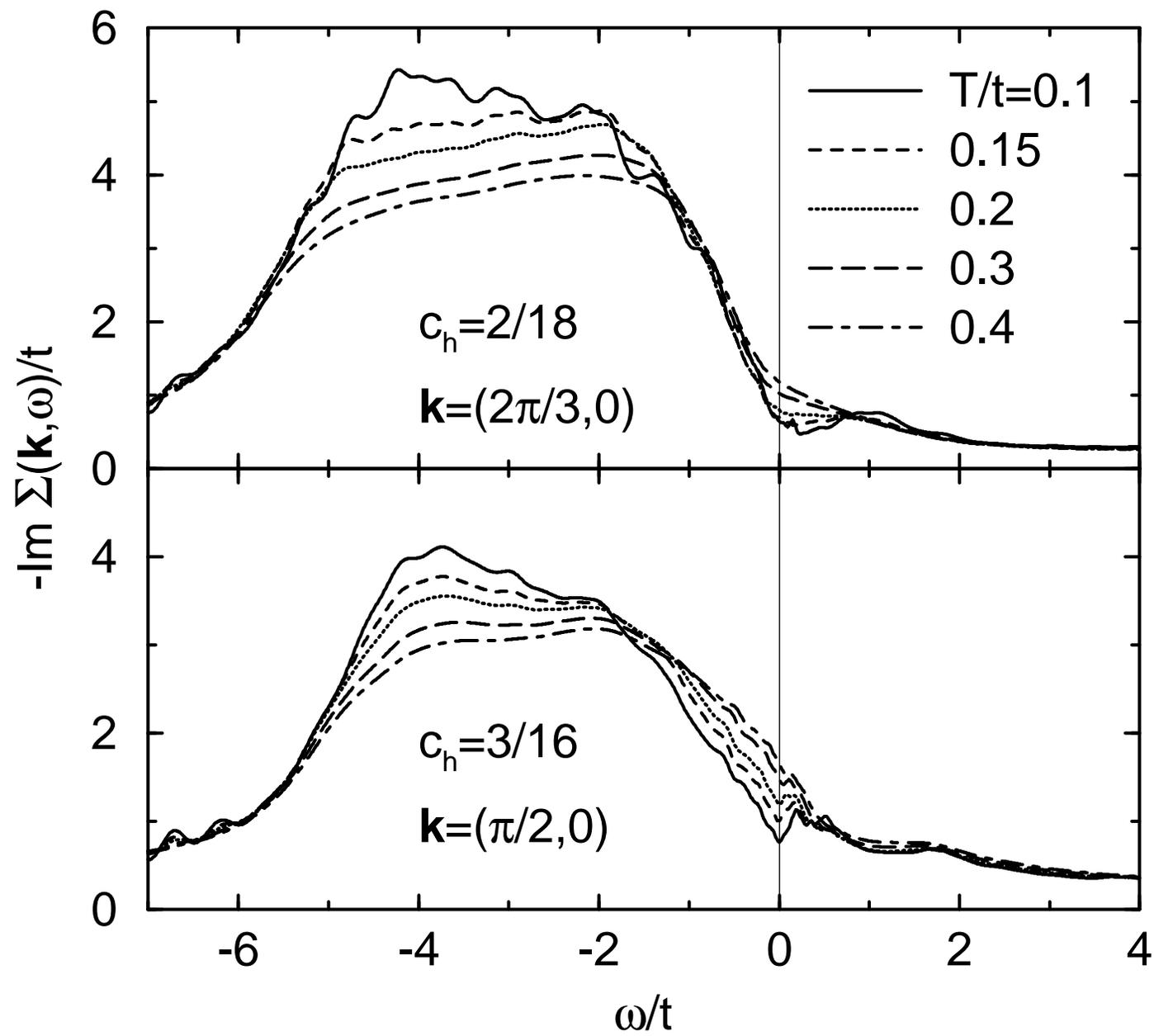
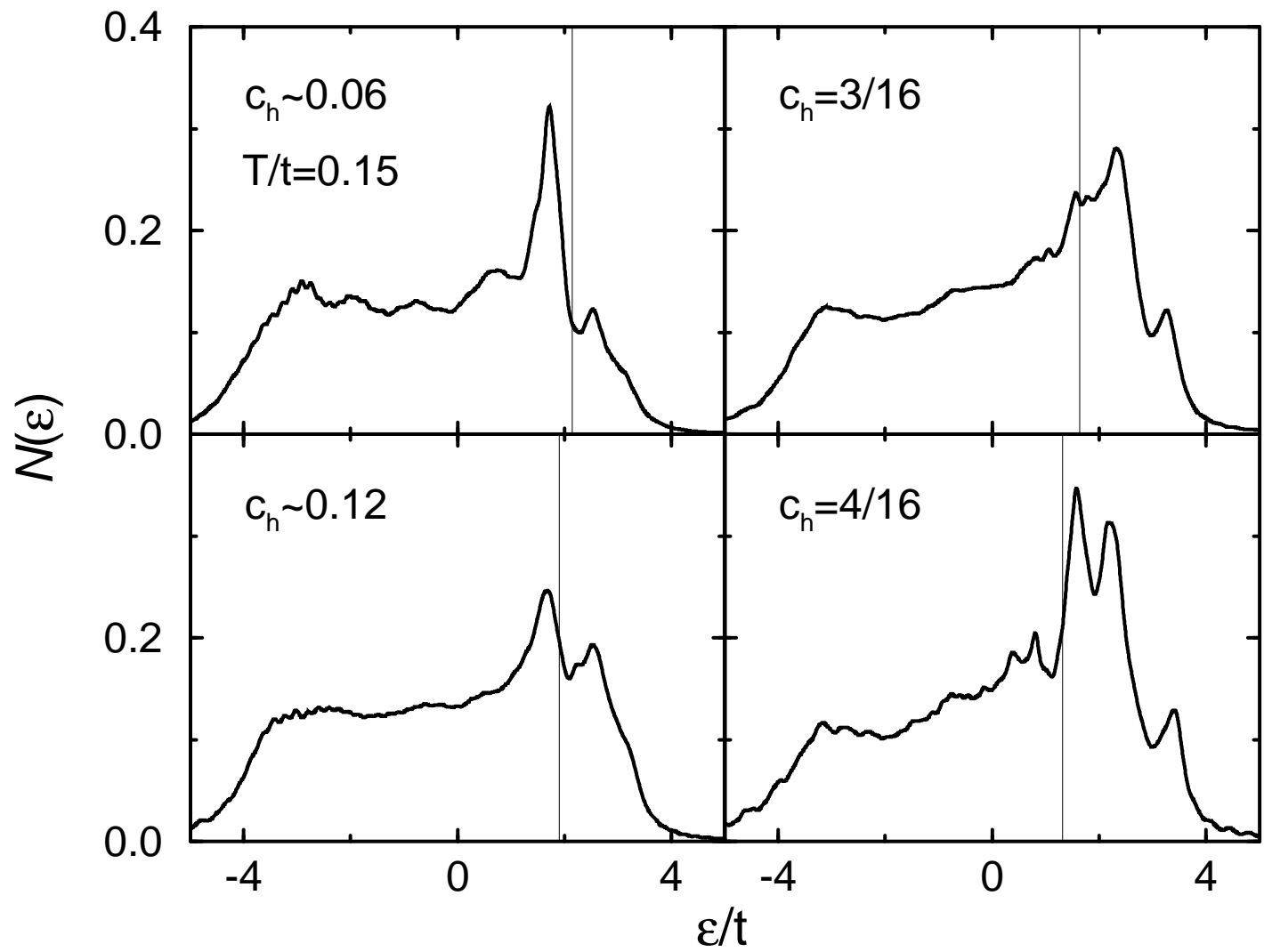
\end{document}